\documentclass[11pt,a4paper]{article}

\usepackage{graphics}
\usepackage{amsmath}
\usepackage{amssymb}
\usepackage{appendix}
\usepackage{multirow}
\usepackage{geometry}
\usepackage{float}
\usepackage{cite}
\usepackage{graphicx}

\usepackage{ifpdf}
\usepackage{color}
\usepackage{hyperref}

\definecolor{nicered}{rgb}{0.7,0.1,0.1}
\definecolor{nicegreen}{rgb}{0.1,0.5,0.1}
\hypersetup{colorlinks,citecolor= nicegreen,linkcolor= nicered}

\def\Aslb{A^b_{\rm sl}}

\def\Adirc{A^c_{\rm dir}}
\def\Adirb{A^b_{\rm dir}}

\newcommand{\beq}{\begin{equation}}
\newcommand{\eeq}{\end{equation}}
\newcommand{\bea}{\begin{eqnarray}}
\newcommand{\eea}{\end{eqnarray}}

\begin{document}

\begin{flushright}
LPT-ORSAY/12-83
\end{flushright}

\vspace{0.3cm}

\begin{center} 
{\bf {\Large On the origin of  
the D{\O} like-sign dimuon charge asymmetry}}

\vspace{0.5cm}

S\'ebastien Descotes-Genon$^{a}$ and Jernej F. Kamenik $^{b,c}$

\vspace{0.2cm}

\emph{$^a$ Laboratoire de Physique 
Th\'eorique, CNRS/Univ. Paris-Sud 11 (UMR 8627),\\
91405 Orsay Cedex, France}\\
\emph{$^b$ J. Stefan Institute, Jamova 39, P. O. Box 3000, 1001 Ljubljana, Slovenia}\\
\emph{$^c$ Department of Physics, University of Ljubljana, Jadranska 19, 1000 Ljubljana, Slovenia}

\end{center}

We reconsider the recent observation by the D\O\, experiment of a sizable like-sign dimuon charge asymmetry, highlighting that it could be affected by CP-violating New Physics contributions not only in $B_d$- and $B_s$-meson mixings, but also in semileptonic decays of $b$ and $c$ quarks producing muons. The D\O\, measurement could be reconciled with the Standard Model expectations for neutral-meson mixings, provided that the CP asymmetry in semileptonic $b$ ($c$) decays reaches 0.3~\% (1\%). Such effects, which lie within the available (rather loose) experimental bounds,  would be clear indications of New Physics and should be investigated experimentally.

\vspace{0.5cm}

%%%%%%%%%%%%%%%%%%%%%%%%%%%%%%%%%%%%%%%%%%%%%%%
%
\section{Introduction}
%
%%%%%%%%%%%%%%%%%%%%%%%%%%%%%%%%%%%%%%%%%%%%%%%

The recent observation by the D\O\, experiment of a sizable like-sign dimuon charge asymmetry~\cite{Abazov:2010hj,Abazov:2010hv,Abazov:2011yk} has attracted tremendous theoretical interest. If interpreted as originating from mixing-induced CP violation in semileptonic $b$ quark decays, the observed value of Ref.~\cite{Abazov:2011yk}
\beq
\Aslb = (-0.787 \pm 0.172 \pm 0.093)\%\,,
\label{eq:ACP}
\eeq
is significantly larger than expected within the standard model (SM) ${\Aslb}^{\rm SM}= (-3.96^{+0.15}_{-0.04})\cdot 10^{-4}$~\cite{Lenz:2012az} (see also Refs.~\cite{ASM}). On the other hand, recent measurements of CP violation in wrong sign semileptonic $B_s$ decays from the LHCb and D\O\, collaborations~\cite{Abazov:2012,LHCb2012}
\beq 
a^s_{\rm sl}=(-1.08\pm 0.72\pm 0.17)\%~[{\rm D\O}]\,,\qquad a^s_{\rm sl}=(-0.24\pm 0.54\pm 0.33)\%~[{\rm LHCb}]\,,
\label{eq:assl}
\eeq
are in agreement with SM expectations. Furthermore, recent precise determinations of the CP asymmetry in the $B_s\to J/\psi \phi$ decay~\cite{phiS}, which are consistent with SM predictions, already severely constrain a possible interpretation of the value in Eq.~\eqref{eq:ACP} in terms of non-standard CP-violating contributions to $B_s$ mixing~\cite{Lenz:2012az}. While the large observed $\Aslb$ could still originate from new CP-violating effects in $B_d$ mixing, this would nonetheless require sizable correlated non-standard contributions also to the absorptive mixing amplitude~\cite{Lenz:2012az} (see also Ref.~\cite{Freytsis:2012ja}).

In view of this intriguing situation, it is important to revisit theoretical assumptions underlying the interpretation of the D\O\, measurement. Since the D\O\, result indicates a significant discrepancy with the SM concerning a tiny CP asymmetry, one should reconsider other possible New Physics (NP) sources which could contribute the observed CP violation, but are generally assumed to be negligible \emph{within the SM}. Indeed one can imagine other sources of CP violation contributing to the dimuon charge asymmetry. Given the sizable charm contributions in both the inclusive muon and same-sign dimuon data samples, an attractive possibility is represented by CP violation in $D^0$ mixing. However, the $D^0$ oscillation probability has been measured and its smallness renders any such contribution negligible. 

On the other hand the D\O\, analysis also assumes that the semileptonic $b$ and $c$ quark decays conserve CP. 
In this letter we explore the consequences of relaxing this assumption. Direct CP violation in semileptonic decays of $b$ or $c$ quarks is presently only poorly constrained experimentally. The related inclusive semileptonic CP asymmetries are expected to be tiny in the SM and so offer interesting venues for contributions of NP. This motivates us to investigate whether the observed dimuon asymmetry in Eq.~\eqref{eq:ACP} could  originate from  CP violation in semileptonic $b$ and $c$ quark decays. 

Previously, a study of the potential impact of CP-violation in semileptonic $b$ decays has been performed in ref.~\cite{BarShalom:2010qr}, finding that the impact on $A^b_{sl}$ was very small in the SM, as well as in a large class of NP models where the main effect comes from the interference between SM tree and NP loop contributions (leading to a generic bound of a few $10^{-6}$). We consider the problem from a different angle, by studying the main ingredients and assumptions behind the D{\O} analysis carefully, including also possible $CP$-violating contributions from semileptonic c decays. We furthermore re-estimate the size of possible SM effects as well as presently experimentally allowed NP contributions to CP asymmetries in semileptonic $b$ and $c$ decays.

The rest of the paper is structured as follows: in Sec.~\ref{sec:exp} we recall the basic elements of the D\O\, analysis needed for the extraction of the CP asymmetries from their inclusive muon and same-sign dimuon data samples. Sec.~\ref{sec:Inerpret} is devoted to the derivation of the relevant observables in presence of CP violation in semileptonic $b$ and $c$ quark decays and the reinterpretation of the D\O\, measurement in terms of the related  CP asymmetries. In Sec.~\ref{sec:bounds} we discuss other existing bounds on  CP violation in these decays, and we give their expectations within the SM and in presence of NP in Sec.~\ref{sec:SM}. Finally, we conclude in Sec.~\ref{sec:End}.

%%%%%%%%%%%%%%%%%%%%%%%%%%%%%%%%%%%%%%%%%%%%%%%
%
\section{Elements of the experimental analysis}
\label{sec:exp}
%
%%%%%%%%%%%%%%%%%%%%%%%%%%%%%%%%%%%%%%%%%%%%%%%

We recall the basic ingredients in the determination of $\Aslb$ from the measurement of the dimuon charge asymmetry according to Ref.~\cite{Abazov:2010hv} and applicable to the most recent update~\cite{Abazov:2011yk} (we will use the values quoted in the latter reference for our numerical analysis). Experimentally, both the like-sign dimuon charge asymmetry $A$ and the inclusive muon charge asymmetry $a$ are measured by counting
\begin{equation}
A=\frac{N^{++}-N^{--}}{N^{++}+N^{--}}\,,\qquad a=\frac{n^{+}-n^{-}}{n^{+}+n^{-}}\,,
\end{equation}
where $N^{++,--}$ ($n^{+,-}$) denote the number of events with two muons (one muon) with a given charge passing the kinematic selections.

There are various detector and material-related processes contributing to these asymmetries. The reconstructed muons are classified into two categories: ``short'' (or S) including muons from weak decays of $b,c,\tau$ and from electromagnetic decays of short-lived mesons ($\phi,\omega,\eta,\rho^0$), and ``long'' (or L) coming from decays of charged kaons and pions as well as from particle misidentification. One can separate the contribution from short muons $a_S$ to the inclusive muon charge asymmetry
\begin{equation}
a=f_S(a_S+\delta)+f_K a_K + f_\pi a_\pi +f_p a_p\,,
\end{equation}
where $\delta$ is the charged asymmetry related to muon detection and identification and to the background $L$ processes. $f_K$ ($f_\pi$, $f_p$) is  the fraction of muons from charged kaon decays (charged pion decays, misidentifications), and $a_K$ ($a_\pi,a_p$) the corresponding charge asymmetry. These quantities are directly measured from experiment.

One has also a similar expression for the dimuon charge asymmetry
\begin{equation}
A=F_{SS} A_S + (F_{bkg}-2F_{LL}) a_S + (2-F_{bkg}) \Delta + F_K A_K + F_\pi A_\pi + F_p A_p\,,
\end{equation}
where $\Delta$ is the detection asymmetry, $F_{K,\pi,p}$ and $A_{K,\pi,p}$ are the fractions and asymmetries related to the various background $L$ processes.

Finally, D\O\, also consider the combination
\begin{equation}
A'=A-\alpha a= F_{SS} A_{S} + (F_{bkg}-2F_{LL}-\alpha f_S) a_S + \ldots\,,
\end{equation}
where $\alpha = 0.959$~\cite{Abazov:2010hv} or $\alpha = 0.89$~\cite{Abazov:2011yk} is tuned to reduce the background uncertainties (denoted by the ellipses), leading to a precise extraction of a linear combination of $a_S$ and $A_s$.

Various processes ($T^\pm_i$) producing $\mu^{\pm}$ can contribute to the asymmetries $a_S$ and $A_S$ with relative weights ($w_i$) as determined from Monte Carlo simulations, see Table~\ref{tab:weights}.\begin{table}
\begin{center}
\begin{tabular}{lll}
  & Process $T^-_i$ ($\to \mu^-$) & Weight  \\
  \hline
1a & $b\to \mu^- X$ & $w_1(1-\chi_0)(1+\Adirb)$ \\
1b & $\bar{b}\to b \to \mu^- X$ & $w_1 \chi_0(1-\Aslb+\Adirb)$ \\
2a & $\bar{b}\to \bar{c}\to \mu^-X$ & $w_2(1-\chi_0)(1+\Adirc)$ \\
2b & $b\to \bar{b} \to \bar{c} \to \mu^-X$ & $w_2\chi_0(1+\Aslb+\Adirc)$ \\
3 & $b\to c\bar{c}q$ or $\bar{b}\to c\bar{c}\bar{q}$ with  $\bar{c}\to \mu^-X$ & $w_3(1+\Adirc)/2$\\
4a & $b\bar b$ with $\eta,\omega,\rho^0,\phi, J/\psi,\psi' \to (\mu^+)\mu^-$ & $w_{4a}/2$ \\
4b & $c\bar c$ with $\eta,\omega,\rho^0,\phi, J/\psi,\psi' \to (\mu^+)\mu^-$ & $w_{4b}/2$ \\
4c & $\eta,\omega,\rho^0,\phi, J/\psi,\psi' \to (\mu^+)\mu^-$ & $w_{4c}/2$ \\
5 & $b\bar{b}c\bar{c}$  with  $\bar{c}\to \mu^-X$& $w_5(1+\Adirc)/2$\\
6 & $c\bar{c}$  with  $\bar{c}\to \mu^-X$ & $w_6(1+\Adirc)/2$\\
\\
  & Process $T^+_i$ ($\to \mu^+$) & Weight \\
  \hline
1a & $\bar{b}\to \mu^+ X$ & $w_1(1-\chi_0)(1-\Adirb)$ \\
1b & $b\to \bar{b} \to \mu^+ X$& $w_1 \chi_0(1+\Aslb-\Adirb)$\\
2a & $b\to c\to \mu^+X$ & $w_2(1-\chi_0)(1-\Adirc)$\\
2b & $\bar{b}\to b \to c \to \mu^+X$& $w_2\chi_0(1-\Aslb-\Adirc)$\\
3 & $b\to c\bar{c}q$ or $\bar{b}\to c\bar{c}\bar{q}$ with $c\to \mu^+X$ & $w_3(1-\Adirc)/2$ \\
4a & $b\bar b$ with $\eta,\omega,\rho^0,\phi, J/\psi,\psi' \to \mu^+(\mu^-)$ & $w_{4a}/2$ \\
4b & $c\bar c$ with $\eta,\omega,\rho^0,\phi, J/\psi,\psi' \to \mu^+(\mu^-)$ & $w_{4b}/2$ \\
4c & $\eta,\omega,\rho^0,\phi, J/\psi,\psi' \to \mu^+(\mu^-)$ & $w_{4c}/2$ \\
5 & $b\bar{b}c\bar{c}$ with $c\to \mu^+X$ & $w_5(1-\Adirc)/2$ \\
6 & $c\bar{c}$  with $c\to \mu^+X$ & $w_6(1-\Adirc)/2$ 
\end{tabular}
\caption{Processes contributing to the inclusive muon and like-sign dimuon samples, adapted from Refs.~\cite{Abazov:2010hv,Abazov:2011yk} to include CP violation in $B_d,B_s$ mixings ($\Aslb$) and in semileptonic decays ($\Adirb,\Adirc$). Numerical values of the weights (normalized to $w_1=1$) used in the analysis are $w_2=0.096\pm 0.012$, $w_3=0.064\pm 0.006$, $w_4=w_{4a}+w_{4b}+w_{4c}=0.021\pm 0.001$, $w_5=0.013\pm 0.002$, $w_6=0.675\pm 0.101$~\cite{Abazov:2011yk}, and the mean $B_{d,s}$ mixing probability $\chi_0=0.147\pm 0.011$~\cite{Abazov:2010hv}  or $\chi_0=0.1259\pm 0.0042$~\cite{Abazov:2011yk}.}
\label{tab:weights}
\end{center}
\end{table}
We start with the probabilities $P_{b(c)}^{\pm}$ for an initial $b(\bar c)$ quark in $b\bar b$ production (prompt $c\bar c$ production, without $b$ hadrons) to produce a $\mu^\pm$, respectively. In the same processes, $\bar P_{b(c)}^{\pm}$ are then the corresponding probabilities to also produce a  $\mu^\pm$ from the accompanying $\bar b( c)$ quark. Finally we have the probabilities $P^\pm_{\rm SLM}=\bar P^\pm_{\rm SLM}$ to produce $\mu^\pm$ from short-lived meson (or SLM: $\eta,\omega,\rho^0,\phi, J/\psi,\psi'$) in events with no $b$ or $c$ hadrons (we assume these processes to conserve CP). We should mention that this table is adapted from Refs.~\cite{Abazov:2010hv,Abazov:2011yk} by distinguishing the CP-conjugate processes in order to include the CP-violating effects discussed in the present letter. 

Assuming all the processes producing muons to be independent\footnote{The correctness of this approximation was  verified numerically using Monte Carlo generated data samples in Ref.~\cite{Abazov:2010hv} for the case of CP violation in $B_{d,s}$ mixing.}, we can now construct both asymmetries as:
\begin{equation} \label{eq:aSP}
a_S=\frac{\sum_{q=b,c,{\rm SLM}}[(P_q^+ + \bar P_q^+)-(P_q^- + \bar P_q^-)]}{\sum_{q=b,c,{SLM}} [( P_q^+ + \bar P_q^+)+(P_q^- + \bar P_q^-)]}\,,
\end{equation}
and
\begin{equation} \label{eq:ASP}
A_S=\frac{\sum_{q=b,c,{\rm SLM}}[(P_q^+ \cdot \bar P_q^+) - (P_q^- \cdot \bar P_q^-)]}{\sum_{q=b,c,{SLM}}[(P_q^+\cdot \bar P_q^+)+(P_q^-\cdot \bar P_q^-)]}\,.
\end{equation}
Let us finally mention that another observable is introduced in Ref.~\cite{Abazov:2011yk} by considering $A'_S$ restricted to a sample of dimuons with a large enough muon impact parameter. We refrain from studying in detail this observable because of our lack of knowledge concerning the experimental inputs and correlations required, but we highlight that it could be analyzed along the same lines as what we present here for $A'_S$.

%%%%%%%%%%%%%%%%%%%%%%%%%%%%%%%%%%%%%%%%%%%%%%%
%
\section{The role of the semileptonic CP asymmetries}
\label{sec:Inerpret}
%
%%%%%%%%%%%%%%%%%%%%%%%%%%%%%%%%%%%%%%%%%%%%%%%

In the absence of  CP violation in semileptonic $b$ and $c$ quark decays,  $a_S$ and $A_S$ can be directly expressed in terms of 
$\Aslb$, a linear combination of the wrong-sign semileptonic flavor specific asymmetries of the $B_{d,s}$ mesons measuring $CP$-violation in their respective mixings~\footnote{As mentioned in the Introduction, possible contributions due to CP violation in $D^0$ mixing are extremely suppressed by the small mixing probability of $D^0$ mesons as measured experimentally.}:
\begin{equation}\label{eq:aslb}
\Aslb=f_d a_{sl}^d+ f_s a_{sl}^s\,, \qquad a_{sl}^q=\frac{\Gamma(\bar{B}_q\to \mu^+X)-\Gamma(B_q\to \mu^-X)}{\Gamma(\bar{B}_q\to \mu^+X)+\Gamma(B_q\to \mu^-X)}\,,
\end{equation}
with $f_d$ and $f_s$ the fractions of $B_d$ and $B_s$ mesons contributing to the asymmetry, which depend on the experimental setting. The values used
by the D\O\, collaboration are either taken from averages at the Tevatron~\cite{Abazov:2010hv}, or at the LEP machines~\cite{Abazov:2011yk}.

In the presence of CP violation in inclusive semileptonic $b$ or $c$ decays, one must define two additional  asymmetries $\Adirb$ and $\Adirc$:
\begin{equation}
\Adirb=\frac{\Gamma(b\to \mu^-X)-\Gamma(\bar{b}\to \mu^+X)}{\Gamma(b\to \mu^-X)+\Gamma(\bar{b}\to \mu^+X)}\,,
\qquad
\Adirc=\frac{\Gamma(\bar{c}\to \mu^-X)-\Gamma(c\to \mu^+X)}{\Gamma(\bar{c}\to \mu^-X)+\Gamma(c\to \mu^+X)}\,.\label{eq:Adirbc}
\end{equation}
Then the most general expressions for the probabilities $P_q^{\pm}, \bar P_{q}^\pm$ read
\begin{subequations}
\begin{eqnarray}
P_b^+&\propto& w_{1b}(1+\Aslb-\Adirb) +w_{2a}(1-\Adirc)+(w_3+w_5) (1-\Adirc)/2 +w_{4a}/2\,,\\
P_b^-&\propto& w_{1a}(1+\Adirb)+w_{2b}(1+\Aslb+\Adirc) +(w_3+w_5) (1+\Adirc)/2 +w_{4a}/2\,,\\
\bar P_b^+&\propto& w_{1a}(1-\Adirb)+w_{2b}(1-\Aslb-\Adirc) +(w_3+w_5) (1-\Adirc)/2 +w_{4a}/2\,,\\
\bar P_b^-&\propto& w_{1b}(1-\Aslb+\Adirb) +w_{2a}(1+\Adirc)+(w_3+w_5) (1+\Adirc)/2 +w_{4a}/2\,,\\
P_c^+&\propto& w_6 (1-\Adirc) +w_{4b}/2\,, \\
\bar P_c^-&\propto& w_6 (1+\Adirc) +w_{4b}/2\,.
\end{eqnarray}
\end{subequations}
Furthermore, since semileptonic charm decay contributions to wrong-sign muons $P_c^-$ and $\bar P_c^+$ are suppressed by the small $D^0$ mixing probability, we have simply 
\begin{equation}
P_c^- = \bar P_c^+ \propto w_{4b}/2\,,
\end{equation}
whereas the short-lived meson decays provide
\begin{equation}
P_{\rm SLM}^\pm=\bar{P}_{\rm SLM}^\pm \propto  w_{4c}/2\,.
\end{equation}
We can then use  Eqs.~(\ref{eq:aSP}) and (\ref{eq:ASP}) to express the three observables related to $S$ muon production
\begin{equation}
a_S(\Aslb,\Adirc,\Adirb), \quad A_S(\Aslb,\Adirc,\Adirb), \quad 
A'_S(\Aslb,\Adirc,\Adirb)=F_{SS}A_S+(F_{bkg}-2F_{LL}-\alpha f_s)a_S\,,
\end{equation}
in terms of the three asymmetries of interest.

Let us emphasize that in principle, one needs to distinguish the weights corresponding to the three different contributions from short-lived mesons $T_{4a,4b,4c}$. Unfortunately, D\O\, does not provide the three probabilities separately. By varying them in the intervals $w_{4c} \in [0,w_4]$, $w_{4b} \in [0,w_4-w_{4c}]$ and $w_{4a} \in [0,w_4-w_{4c}-w_{4b}]$, we have checked however that the associated additional uncertainty in relating $\Aslb$, $\Adirb$ and $\Adirc$ to $a_S$ and $A_S$ is negligible (for definiteness, we present the results corresponding to $w_{4b}=w_{4c}=0$ in the following). In the limiting case where $\Adirb=\Adirc=0$ and $w_{4b}=w_{4c}=0$, the resulting expressions
coincide with the corresponding expressions in Ref.~\cite{Abazov:2011yk}.

Unfortunately, lacking complete information on the correlations among the various inputs entering the D\O\, measurement, we cannot extract the values of $\Adirb$ and/or $\Adirc$ from the measurements of $a$, $A$ (and $A'$) in a trustable way. However, using $a_S,A_S$ and $A'_S$,
we are in the position to at least estimate the values of $\Adirc$ or $\Adirb$ assuming $\Aslb$ as predicted in the SM~\footnote{We consider the SM result from Ref.~\cite{Lenz:2012az}, which is larger in size than the SM result quoted in Refs.~\cite{Abazov:2010hv,Abazov:2011yk}, and thus our results can be considered more conservative concerning the required size of $CP$-violation in semileptonic $b$ or $c$ decays.}.
We do this by looking for solutions of the following equations 
\begin{subequations}
\begin{align}
\label{eq:eq1}
a_S({\Aslb}^{\rm SM},\Adirc,0)&= a_S({\Aslb}_{,a},0,0)\,, \\
 A_S({\Aslb}^{\rm SM},\Adirc,0)& = A_S({\Aslb}_{,A},0,0)
\,, \\
 A'_S({\Aslb}^{\rm SM},\Adirc,0)& =  A_S({\Aslb}_{,A'},0,0)\,,
\label{eq:eq3}
\end{align}
\end{subequations}
and similarly for the case of nonzero $ \Adirb$ instead of  $ \Adirc$. Here ${\Aslb}_{,a},{\Aslb}_{,A},{\Aslb}_{,A'}$ are values of ${\Aslb}$ as extracted from $a$, $A$ or $A'$ in Ref.~\cite{Abazov:2011yk}. The results are collected in Table~\ref{tab:Acb}. 
\begin{table}
\begin{center}
\begin{tabular}{c|ccc}
CP asymmetry (\%)  &  $a$ & $A$ & $A'$ \\     
\hline\hline
$\Aslb\ (\Adirc=0,\Adirb=0)$~\cite{Abazov:2011yk} & $-1.04\pm 1.30\pm 2.31$ & $-0.808 \pm 0.202 \pm 0.222$ & $-0.787\pm 0.172 \pm 0.093$
 \\
$\Adirb\ (\Aslb={\Aslb}^{\rm SM},\Adirc=0)$   & 
  $0.11\pm 0.15\pm 0.28$ & $0.26\pm 0.07\pm 0.07$ & $0.28\pm 0.06\pm 0.04$\\
$\Adirc\ (\Aslb={\Aslb}^{\rm SM},\Adirb=0)$  & 
   $0.13\pm 0.17\pm 0.31$ & $0.69\pm 0.18\pm 0.21$ & $0.93\pm 0.21\pm 0.20$\\
 \end{tabular}
\caption{CP asymmetries (in \%) needed in meson mixing (first row) or semileptonic decays (second and third rows) in order to obtain the measured values of $a,A,A'$ from Ref.~\cite{Abazov:2011yk}.}
\label{tab:Acb}
\end{center}
\end{table}
The uncertainties quoted are dominated by those coming from ${\Aslb}_{,a},{\Aslb}_{,A},{\Aslb}_{,A'}$. These implicitly include the uncertainties (and correlations) from the other parameters entering the D\O~analysis.  
We furthermore include explicit contributions from additional input parameters entering equations~\eqref{eq:eq1}-\eqref{eq:eq3} to the systematical uncertainties (without the proper knowledge of their correlations with ${\Aslb}_{,a},{\Aslb}_{,A},{\Aslb}_{,A'}$, this can only be considered a very rough estimate of the potential size of their effects). 
We note that these additional contributions only significantly affect the extraction of $\Adirc$ from $A'$, where they almost double the systematic error budget.  On the other hand, the uncertainties coming from the SM prediction for ${\Aslb}^{\rm }$ are completely subleading and thus not quoted. 
Finally, we have checked that consistent results (but with slightly larger errors) are obtained when using inputs and asymmetry measurements from the previous D\O\, analysis~\cite{Abazov:2010hj,Abazov:2010hv}.

One notices that CP asymmetries in inclusive semileptonic $b$ $(c)$ decays below the $0.3\% (1\%)$ level are required to explain the D\O\, measurement. We also note that the values of $\Adirb$ as extracted from $a$ and $A$ are well consistent at $0.4\sigma$ level (assuming Gaussian uncertainties and in absence of correlations), while there is a slight $1.2\sigma$ difference between the values of $\Adirc$ extracted the same way. Although the values extracted from $A'$ are even bigger, they are expected to be highly correlated with the ones from the other two observables, and we do not attempt to assign a statistical meaning to these differences. 

One can imagine that both $\Adirb$ and $\Adirc$ may differ from zero. It is then useful to compute the dependence of $a_S$, $A_S$ and $A'_S$ on the three CP asymmetries (to first order), yielding:
\begin{subequations}
\begin{eqnarray}
a_S &=& \Aslb(0.061\pm0.004)+\Adirb(-0.535\pm0.028)+\Adirc(-0.454\pm0.028)\,,\\
A_S &=&\Aslb(0.474\pm0.023)+\Adirb(-1.421\pm0.024)+\Adirc(-0.527\pm0.025)\,,\\
A'_S &=&\Aslb(0.312\pm 0.023)+\Adirb(-0.849\pm 0.061)+\Adirc(-0.250\pm 0.038)\,,
\end{eqnarray}
\end{subequations}
where the relevant systematical and statistical uncertainties have been combined in quadrature. We see that the observables $a_S, A_S,A'_S$ exhibit similar sensitivities to the three types of CP violation. This clearly indicates that the interpretation of these quantities in terms of neutral-meson mixing requires a further check of the absence of CP violation in decays at a similar level to the uncertainties quoted for $\Aslb$.

\begin{figure}
\begin{center}
\includegraphics[width=8cm]{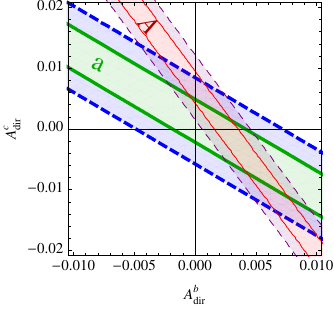}
\end{center}
\caption{The $\Adirc$  and $\Adirb$  semileptonic decay asymmetries needed in order to reproduce the measured values of inclusive semileptonic and same-sign dimuon asymmetries $a$ (in thick contours) and $A$ (in thin contours) (results for $A'$ closely resemble those for $A$) using the SM predicted value for the inclusive wrong-sign semileptonic asymmetry $\Aslb$ (the weights and parameters of Ref.~\cite{Abazov:2011yk} have been used). The $1\sigma(2\sigma)$ bands are marked with full (dashed) contours.}
\label{fig:Acb}
\end{figure}

Assuming the SM value of $\Aslb$, the experimental values of $a_S$ and $A_S$ set constraints in the $(\Adirb,\Adirc)$ plane, as illustrated in Fig.~\ref{fig:Acb}.  One can see that the sensitivity of the observables to $\Adirb$ is larger than that to $\Adirc$, explaining that a larger asymmetry in charm is required to reproduce the D\O\, value for the dimuon asymmetry. Since our analysis does not include all the relevant correlations, we do not attempt at combining the two constraints statistically, even though this could be done easily by the D\O\, collaboration.

%%%%%%%%%%%%%%%%%%%%%%%%%%%%%%%%%%%%%%%%%%%%%%%
%
\section{Other experimental constraints on Direct $CP$ violation in semileptonic $b$ and $c$ decays}
\label{sec:bounds}
%
%%%%%%%%%%%%%%%%%%%%%%%%%%%%%%%%%%%%%%%%%%%%%%%

At present and to the best of our knowledge, there exist no direct experimental bound on the $CP$-violation in semileptonic decays of either $b$ or $c$ quarks~\cite{Nakamura:2010zzi,HFAG}.  We can nonetheless form a poor man's estimate of the current constraints by considering the uncertainties on the CP averaged (semi)leptonic branching fractions (estimating in this way the amount of $CP$-violation in semileptonic decays that could be hidden by the systematics of the measurement).  In particular, among the various measured semileptonic $D^+$, $D^0$ and $D_s$ decays, the CP averaged branching fraction of $D^+ \to \bar K^0 \mu^+ \nu_\mu$ (which accounts for more than half the inclusive semileptonic $D^+$ branching fraction) is currently most precisely known with $6.5\%$ relative uncertainty~\cite{Ablikim:2006bw}. Based on this and considering also the experimental bound on the CP asymmetries in leptonic $D^\pm \to \mu^\pm \nu$ decays~\cite{Eisenstein:2008aa} we observe that at the current level of uncertainty, $CP$-violation in semileptonic $D$ decays at or below 6\% is still viable. 

Concerning  leptonic and semileptonic decays of $B$ mesons, we find that the current level of uncertainty of the CP averaged inclusive $B \to X_c \ell \nu$ branching fractions~\cite{Nakamura:2010zzi} in principle still allows for $CP$ violation in semileptonic $B$ decays at the level of around 3\%.
However in this case, additional constraints can be derived using the measured flavor specific semileptonic charge asymmetries defined in Eq.~\eqref{eq:aslb}~\cite{HFAG}, which yields for the $B_d$ meson 
 \begin{equation}
a^d_{sl}=(-0.05\pm 0.56)\%\,,
\end{equation}
whereas the situation for $a^s_{sl}$ is recalled in Eq.~(\ref{eq:assl}). These measurements are generally based
on specific charmed decay modes associated with muons, so that they can be affected by CP violation in semileptonic $b$ decays, but not in $c$ decays, contrary to the D\O\, measurement.

In reinterpreting these results in terms of CP-violating asymmetries in inclusive semileptonic $b$ decays at D\O\,, we first identify $|a^d_{sl}| \gtrsim |A_{\rm dir}^{B_d}|$ and $|a^s_{sl}| \gtrsim  |A_{\rm dir}^{B_s}|$, where the semileptonic asymmetries $A_{\rm dir}^{B_q}$ are defined as in Eq.~\eqref{eq:Adirbc}, but now refer to the relevant decaying $\bar B_q$ mesons. 
Next we need to sum over the relative production fractions of the various $b$ hadrons contributing to the semileptonic event samples: 
\beq
\Adirb = f({B_u}) A_{\rm dir}^{B_u} + f({B_d}) A_{\rm dir}^{B_d} + f({B_s}) A_{\rm dir}^{B_s} 
%+ f_{b-\rm baryons} A^{b-\rm baryons} 
+ \ldots\,,
\eeq
where the ellipses denote neglected smaller $b$ hadronic state contributions. As a first approximation for $f({B_q})$, we neglect different lifetime effects and use the measured unbiased $b$-hadron fractions as measured by various experiments at high energies $f({B_d}) = f({B_u}) = 0.401\pm0.007$ and $f({B_s})=0.107\pm 0.005$
%$f_{b-\rm baryon} = 0.091\pm 0.015$
~\cite{HFAG}. 
Neglecting possible correlations between the various inputs and combining all uncertainties in quadrature we obtain a bound of $|\Adirb| \lesssim 1.2\%$\, which is safely above what is required for explaining the D\O\, result.

%%%%%%%%%%%%%%%%%%%%%%%%%%%%%%%%%%%%%%%%%%%%%%%
%
\section{SM and NP Expectations for $\Adirb$ and $\Adirc$}
\label{sec:SM}
%
%%%%%%%%%%%%%%%%%%%%%%%%%%%%%%%%%%%%%%%%%%%%%%%

Finally let us briefly comment on the expected size of $\Adirb$ and $\Adirc$ within the SM and in presence of NP.  
Direct CP violation in decays requires the presence of (at least) two interfering decay amplitudes  (we will denote them as $\mathcal A_T \equiv |\mathcal A_T| \exp i (\phi_T+\delta_T)$ and $\mathcal A_L \equiv |\mathcal A_L| \exp i (\phi_L + \delta_L) $) with different weak ($\phi_{T,L}$) and strong ($\delta_{T,L}$) phases. Denoting $\mathcal R \equiv |\mathcal A_L / \mathcal A_T|$, $\Delta \phi = \phi_L-\phi_T$ and $\Delta \delta = \delta_L-\delta_T$, the related  CP asymmetry can then be written as
\beq
\label{eq:Adir}
A_{\rm dir} = \frac{2 \mathcal R \sin \Delta \delta \sin \Delta \phi}{1+ 2 \mathcal R \cos \Delta \delta \cos \Delta \phi + \mathcal R^2}\,.
\eeq

The dominant tree-level SM contributions to semileptonic transitions (below the $W$ scale) are described by the relevant effective weak Hamiltonian
\beq
\mathcal H_{\rm eff}^{sl} = \frac{4 G_F}{\sqrt 2} \sum_{U=u,c}\sum_{D=d,s,b} V^*_{UD} [\bar U \gamma_\mu (1-\gamma_5) D] [\bar \ell \gamma^\mu (1-\gamma_5) \nu_\ell] + \rm h.c.\,.
\eeq
leading to a very simple amplitude $\mathcal A_T$.
 The required additional amplitudes $\mathcal A_L$ can be generated at one-loop via time ordered correlators of $\mathcal H_{\rm eff}^{sl}$ with the effective weak Hamiltonian $\mathcal H_{\rm eff}^{nl}$ describing non-leptonic $b\to c \bar c d$ and  $b\to c \bar c s$ decays $\int d^4 x T\{ \mathcal H_{\rm eff}^{nl}(0), \mathcal H_{\rm eff}^{sl} (x) \}$. The presence of on-shell quarks in the loop provides a source for the strong phase difference, while the weak phase difference is encoded in the relevant CKM matrix elements. However, being higher order in $G_F \sim 1/v_{EW}^2$ these effects are expected to be severely suppressed leading to $\mathcal R \ll 1$. Expanding Eq.~\eqref{eq:Adir} to linear order in $\mathcal R$, assuming $\sin\Delta \delta=O(1)$ and using na\"ive dimensional analysis, we can estimate the resulting SM contributions to $\Adirb$ and $\Adirc$ coming from the interference between tree level and one loop as
\begin{subequations}
 \begin{align}\label{eq:AdirbSM}
 {\Adirb}^{\rm SM} &\sim 2 \frac{1}{4\pi}\left(\frac{m_b}{v_{EW}}\right)^2 {\rm Im} \left( \frac{V_{tb} V^*_{ts} V_{cs}}{V_{cb}} \right) \sim 10^{-6}\,, \\
  {\Adirc}^{\rm SM} & \sim 2 \frac{1}{4\pi} \left(\frac{m_c}{ v_{EW}}\right)^2 {\rm Im} \left( \frac{V_{cb} V^*_{ub} V_{us}}{V_{cs}} \right) \sim 10^{-10}\,.
 \end{align}
 \end{subequations}
We observe that the sizes of $\Adirb$ or $\Adirc$ required to accommodate the D\O\, result are orders of magnitude larger than the above SM expectations for these quantities and their confirmation would thus constitute a clear indication of NP. 

A more elaborate discussion of the size of ${\Adirb}^{\rm SM}$ was performed in ref.~\cite{BarShalom:2010qr}, with a final result suppressed by another three orders of magnitude compared to the na\"ive estimate (21a) due to the more careful consideration of intermediate state phase-space effects and the evaluation of the relevant  penguin operator Wilson coefficients in $\mathcal H_{\rm eff}^{nl}$.

In the same reference, a generic bound for NP contributions was presented for a large class of NP models where the main effect comes from the interference between SM tree and NP loop contributions (leading to a generic bound of a few $10^{-6}$), and the potential contribution to $\Adirb$ arising in the framework of a left-right symmetric model was studied. The experimental constraints on the right-handed charged currents led to values for the asymmetry of a similar size to the na\"ive estimate in eq.~(\ref{eq:AdirbSM}) (around $10^{-7}$), suggesting that $\Adirb \sim \mathcal O(10^{-3})$ would be difficult to accommodate with simple NP models.

While an explicit NP model construction reproducing the required effects is clearly beyond the scope of the present study, we briefly mention two possibilities of circumventing the generic bound of ref.~\cite{BarShalom:2010qr} by selecting cases where its underlying assumptions (NP enters only via charged current loops) are not fulfilled. The first example involves introducing a light neutral $Z'$ coupling only to quarks (with strength $g^{q,q'}_{Z'}$ naturally carrying a CP-odd phase), and charged under a non-abelian family symmetry so that dangerous tree-level $\Delta F=2$ FCNCs are forbidden. Such neutral vector bosons are also only weakly constrained by direct searches and electroweak precision tests (c.f.~\cite{Grinstein:2011dz}). At the same time an effective $\bar c b \bar d u$ interaction can be generated, suppressed by a mass scale ($m_{Z'}/\sqrt{|g^{bd}_{Z'}g^{uc*}_{Z'}|}$) comparable or even below the SM weak scale. The resulting $\Adirb$ can be correspondingly enhanced compared to the SM estimate by a factor $(v_{EW}/m_{Z'})^2 {\rm Im}(g^{bd}_{Z'}g^{uc*}_{Z'} /V_{ub} V_{cd})$, thus in principle circumventing the bound in ref.~\cite{BarShalom:2010qr}. 

The second possibility exists through final states involving light invisible particles mimicking the missing energy signature of the SM neutrinos in semileptonic decays (see ref.~\cite{invisibles} for recent work along these lines). For example one can consider lepton number (and CP) violating interactions of the form $\bar c b \bar \ell \chi_i$, where $\chi_{i}$  ($i=1,2$) are new light neutral fermions. If these fermions are very short lived (such that their decay widths are comparable to their masses) and decay to a common (invisible) final state, sizeable CP asymmetries can be generated from interferences of amplitudes with intermediate $\chi_1$ and $\chi_2$ (see~\cite{resonance} for previous discussions of this mechanism). Such incoherent effects, with new intermediate states and interactions, could thus in principle provide another way to circumvent the generic bound of ref.~\cite{BarShalom:2010qr}.

%%%%%%%%%%%%%%%%%%%%%%%%%%%%%%%%%%%%%%%%%%%%%%%
%
\section{Conclusions}
\label{sec:End}
%
%%%%%%%%%%%%%%%%%%%%%%%%%%%%%%%%%%%%%%%%%%%%%%%

We have reconsidered the recent measurement made by the D\O\, collaboration of a like-sign dimuon asymmetry~\cite{Abazov:2010hj,Abazov:2010hv,Abazov:2011yk}. Since this measurement disagrees significantly with the SM prediction for CP violation in $B_d$ and $B_s$-meson mixing, we have reassessed some of the underlying assumptions of the analysis, allowing for NP effects violating CP not only in mixing, but also in decays.

We have shown that the D\O\, result can be made compatible with the SM expectations for CP violation in $B_{d,s}$ mixing, provided that non-SM contributions are introduced either in
\begin{itemize}
\item CP violation in $b$ semileptonic decays ($\Adirb\simeq 0.3\%$), which is currently allowed experimentally, though close to the uncertainties quoted for the individual semileptonic $B_d$ and $B_s$ asymmetries, and difficult to accommodate with simple NP models.
\item CP violation in $c$ semileptonic decays ($\Adirc\simeq 1\%$), which is currently allowed experimentally.
\end{itemize}
As indicated in Fig.~\ref{fig:Acb}, one could also consider the presence of both effects. We discussed
briefly the size of these effects in the SM, and showed that they are much smaller than what is required to explain the D\O\, result.

%At this stage, the possibility of a CP violation in semileptonic charm decays (at the percent level) is actually quite intriguing and one could speculate about the connection of such effect with the significant CP violation observed in hadronic charmed decays~\cite{Aaij:2011in,Aaltonen:2011se}. 

In particular, a CP-violating contribution to charm semileptonic decays would allow the D\O\, measurement of the like-sign dimuon asymmetry to differ from the SM value, but it would not affect the measurements of $a^q_{ sl}$ based on specific decay channels like $B_s\to D_s\mu X$. 

Our analysis is obviously very na\"ive as far as experimental uncertainties and correlations are concerned. The numbers provided in this short note are only indicative, but we firmly hope that they will incite experimentalists to revisit the D\O\, analysis and related studies to include and constrain CP violation in semileptonic $b$ and $c$ decays. Such cross-checks would be particularly useful to improve our understanding of neutral-meson mixing and its potential for NP searches.

\section*{Acknowledgments}
We thank Michael Gronau for pointing out a mistake in our original evaluation of the SM contributions to the direct semileptonic CP asymmetries in B decays. We also thank Guennadi Borissov, J\'er\^ome Charles, Bruce Hoeneisen, Bostjan Golob and Karim Trabelsi for useful discussions. Work partially supported by PHC-PROTEUS 2012, Project 26807QH and by the Slovenian Research Agency.

\end{document}